\documentclass[12pt]{iopart}

\long\def\symbolfootnote[#1]#2{\begingroup%
\def\thefootnote{\fnsymbol{footnote}}\footnote[#1]{#2}\endgroup}

\newcommand{\be}{\begin{equation}}
\newcommand{\ee}{\end{equation}}
\newcommand{\bea}{\begin{eqnarray}}
\newcommand{\eea}{\end{eqnarray}}
\newcommand{\ba}{\hspace*{-5pt}\begin{array}}
\newcommand{\ea}{\end{array}}
\newcommand{\p}{\partial}
\newcommand{\ds}{\displaystyle}

\usepackage{iopams}
\usepackage{graphicx}
\usepackage{dcolumn}
\usepackage{bm}
\let\p\partial

\begin{document}


\title[Ordering of parameters in the water wave problem]
{Ordering of two small parameters in the shallow water wave problem\symbolfootnote[0]{\hspace*{-3ex}\copyright\ 2013 IOP Publishing Ltd\\[3mm]
{\em J. Phys. A: Math. Theor.} 2013, to appear\\[3mm]
This is an author-created, un-copyedited version of an article accepted for publication in {\em J. Phys. A: Math. Theor.} IOP Publishing Ltd is not responsible for any errors or omissions in this version of the manuscript or any version derived from it.}}

\author{Georgy I. Burde$^1$ and Artur Sergyeyev$^2$}

\address{$^1$ Jacob Blaustein Institutes for
Desert Research, Ben-Gurion University, Sede-Boker Campus, 84990
Israel}

\address{$^2$ Mathematical Institute, Silesian University
in Opava, Na Rybn\'\i{}\v{c}ku 1, 74601
Opava, Czech Republic}

\ead{\mailto{georg@bgu.ac.il} \normalfont{and} \mailto{Artur.Sergyeyev@math.slu.cz}}

\begin{abstract}

The classical problem of irrotational long waves on the surface of
a shallow layer of an ideal fluid moving under the influence of
gravity as well as surface tension is considered. A systematic
procedure for deriving an equation for surface elevation for a
prescribed relation between the orders of the two expansion
parameters, the amplitude parameter $\alpha$ and the long
wavelength (or shallowness) parameter $\beta$, is developed.
Unlike the heuristic approaches found in the literature, when
modifications are made in the equation for surface elevation
itself, the procedure starts from the consistently truncated
asymptotic expansions for unidirectional waves, a counterpart of
the Boussinesq system of equations for the surface elevation and
the bottom velocity, from which the leading order
and higher order equations for the surface elevation can be
obtained by iterations. The relations between the orders of the
two small parameters are taken in the form $\beta=O(\alpha^n)$ and
$\alpha=O(\beta^m)$ with $n$ and $m$ specified to some important
particular cases. The analysis shows, in particular, that some
evolution equations, proposed before as model equations in other
physical contexts (like the Gardner equation, the modified KdV
equation, and the so-called 5th-order KdV equation), can emerge as
the leading order equations in the asymptotic expansion for the
unidirectional water waves, on equal footing with the KdV
equation. The results related to the higher orders of
approximation provide a set of consistent higher order model
equations for unidirectional water waves which replace the KdV
equation with higher-order corrections in the case of non-standard
ordering when the parameters $\alpha$ and $\beta$ are not of the
same order of magnitude. The shortcomings of certain models used
in the literature become apparent as a result of the subsequent
analysis. It is also shown that various model equations obtained by assuming
a prescribed relation $\beta=O(\alpha^n)$ between the orders of
the two small parameters can be equivalently treated as obtained
by applying transformations of variables which scale out the
parameter $\beta$ in favor of $\alpha$. It allows us to consider
the nonlinearity-dispersion balance, epitomized by the soliton
equations, as existing for any $\beta$, provided that
$\alpha\rightarrow 0$, but leads to a prescription, in asymptotic
terms, of the region of time and space where the equations are
valid and so the corresponding dynamics is expected to occur.\vspace{-7mm}
\looseness=-1
\end{abstract}


\bigskip


\section{Introduction}

The behavior of surface gravity waves on shallow water has been a
subject of intense research. In particular, the famous
Korteweg--de Vries (KdV) equation, which is the prototypical
example of an exactly solvable soliton equation, was first
introduced as a unidirectional nonlinear wave equation obtained
via asymptotic expansion around simple wave motion of the Euler
equations for shallow water.

The system of equations describing the long, small-amplitude wave
motion in shallow water with a free surface
\cite{whitham}--\cite{johnson1} involves two \textit{independent}
small parameters: $\alpha$, which measures the ratio of wave
amplitude to undisturbed fluid depth, and $\beta$, which measures
the square of the ratio of fluid depth to wave length, and no
relationship between orders of magnitude of $\alpha$ and $\beta$
follows from the statement of the problem. The KdV equation
\begin{equation}
\eta_t+\eta_x+\frac32\alpha\eta\eta_x+\frac{1}{6}\beta\eta_{3x}=0
\label{01}
\end{equation}
emerges at first order (in both parameters $\alpha$ and $\beta$)
in the asymptotic expansion as an equation for the surface
elevation $\eta$ associated with the right-moving wave. The
derivation assumes (sometimes tacitly) that $\beta=O(\alpha)$.
It is evident that in the case, when $\alpha$ and $\beta$ differ
in their orders of magnitude, the leading order equation
maintaining the balance between linear dispersion and nonlinear
steepening, which is the primary physical mechanism for the
propagation of solitary shallow water waves, should change its
form. The same holds true for the equations which (like the
higher-order KdV equations) address higher order effects.

A heuristic approach to deriving model equations for
unidirectional water waves is frequently used when some additional
terms are included into the equation for the surface elevation
based on relations between the orders of parameters. However, this
may lead to inconsistencies.
For example, the assumption $\alpha\ge\beta>\alpha^2$ is made in
\cite{gottwald0} while the terms involving $\alpha$, $\beta$,
$\alpha^2$, $\alpha \beta$ and $\beta^2$ are kept and,
accordingly, the terms involving $\alpha^3$, $\alpha^2\beta$,
$\alpha\beta^2$ and $\beta^3$ are neglected. It is readily seen
that the relation $\beta=\alpha^r$ with $3/2\le r\le 2$ which
satisfies the above inequality is in conflict with the truncation
made: the neglected term $\sim \alpha^3$ is as important as the
retained term $\sim \beta^2$. In  \cite{greki}, based on an
inequality $O(\beta)<O(\alpha)$, the truncation is made such that
the terms involving $\alpha$, $\beta$, $\alpha^2$, $\alpha \beta$,
$\alpha^3$ and $\alpha^2\beta$ are kept and the terms involving
$\alpha^4$, $\alpha^3\beta$ and $\beta^2$ are neglected. However,
this choice of truncation is questionable since there exists no
relationship between the orders of $\beta$ and $\alpha$ of the
form $\beta =O(\alpha^r)$ (or $\alpha =O(\beta^r)$) for which such
a truncation is consistent. Indeed, assuming $\beta =O(\alpha^r)$
with $r>1$ (which is compatible with $O(\beta)<O(\alpha)$), one
can see that the two requirements $\alpha^4<\alpha^2\beta$ and
$\beta^2<\alpha^2\beta$ lead to conflicting results: $r<2$ and
$r>2$ respectively. Thus, such a heuristic approach does not
provide a reliable way to determine even a form of the equation
for surface elevation and, what is more, it does not allow
determining coefficients of the equation. It is well known that
the solution properties may strongly depend on the relations
between the coefficients -- the higher order KdV equations can be
mentioned in this respect (see e.g.\
\cite{kichenassamy}--\cite{burde2} and references therein).
\looseness=-1

In general, to arrive at a consistent model equation for water
waves, the ordering of terms should be made in the original asymptotic
expansion for unidirectional water waves based on a prescribed
relationship between orders of magnitude of $\alpha$ and $\beta$.
Then a consistent truncation of the expansion can be made and the
related leading order and higher order evolution equations can be
defined. In the present paper such a procedure for deriving an
equation for surface elevation for a prescribed relation between the
orders of the two expansion parameters $\alpha$ and $\beta$ is
developed. It makes possible a systematic study of different particular
cases and corresponding leading order and higher order equations.
The following special cases are considered: $\beta=O(\alpha^2)$,
$\alpha=O(\beta^2)$, $\beta=O(\alpha^3)$ and $\alpha=O(\beta^3)$.
The analysis is aimed at deriving an equation for the surface
elevation having a form of an evolution equation; therefore,
equations which, like the Benjamin--Bona--Mahoney (BBM) equation
\cite{bbm}, contain the time derivatives in the higher order terms
are excluded from consideration. The results of the analysis show,
in particular, that some evolution equations proposed before as
model equations in other physical contexts can play the role of a
model equation at the leading order of the asymptotic expansion
for the unidirectional water waves on equal footing with the KdV
equation. Some of these equations, both integrable and
non-integrable, are known to have a rich structure of solitary
wave solutions which differ in their properties from the KdV
solitons. Thus, the leading order soliton dynamics in the
unidirectional water wave problem can differ from the one
described by the KdV equation. The equations obtained in the
higher orders of approximation, in general, also differ from the
KdV equation with higher order corrections. It is worth noticing
that the above differences from the standard model are {\em not}
due to taking the surface tension into account. New equations and
dynamics arise even in the classical formulation, when capillary
effects are neglected, if the ordering is non-standard ($\beta$
and $\alpha$ are not of the same order of magnitude). Including
surface tension in general does not alter the structure of the
leading order and higher order equations, only some specific
cases, like the case $\tau=1/3$ of the standard analysis, should
be considered separately.

The paper is organized as follows.
In Section 2 following the Introduction, we present the statement of
the problem, the basic equations and the outline of the procedure. The
main ideas of the analysis are described in more detail in
Section 3 where the procedure is presented for the best studied case
of $\beta=O(\alpha)$. The cases when the relation
$\beta=O(\alpha)$ does not hold are studied in the subsequent Section 4.
In the said section the analysis is restricted to the pure gravity
waves in order to better explain the main points and also
to demonstrate that the differences from the standard model are not
due to taking the surface tension into account. The results
for the gravity-capillary waves are listed in the Appendix.
In Section 5, the concluding remarks are given and an alternative
interpretation of the results, based on a transformation of
variables which scales out the parameter $\beta$, in favor of
$\alpha$, is discussed.

\section{Outline of the procedure}

Consider the standard system of equations describing the
two-dimensional irrotational wave motion of an inviscid
incompressible fluid in a channel with the flat horizontal rigid
bottom and the free surface under the influence of gravity as well
as surface tension. After an appropriate choice of non-dimensional
variables, the equations of motion and boundary conditions can be
reduced to the system written in terms of the velocity potential
$\phi(x,y,t)$ and the surface elevation $\eta(x,t)$, see e.g.\
\cite{whitham}:
\begin{eqnarray} &\beta \phi_{xx}+\phi_{yy}=0,\qquad
0\leq y\leq 1+\alpha
\eta\label{6} &\\
&\phi_{y}=0,\qquad y=0\label{7} &\\
&\eta_{t}+\alpha \phi_{x}\eta_{x}
-\frac{1}{\beta}\phi_{y}=0,\qquad y=1+\alpha \eta\label{8} &\\
&\phi_{t}+\frac{1}{2}\alpha \phi_{x}^2
+\frac{1}{2}\frac{\alpha}{\beta}\phi_{y}^2+\eta -\tau \beta
\frac{\eta_{xx}}{\left (1+ \alpha^2\beta \eta_{x}^2\right)^{3/2}}
=0,\qquad y=1+\alpha \eta\label{9}&
\end{eqnarray}
where $t$ is time and $x,y$ are respectively horizontal and
vertical coordinates, with $y=0$ being the bottom. The
non-dimensional variables are defined as follows (after
non-dimensionalizing, the tildes have been omitted):
\begin{equation} \tilde x=\frac{x}{L},\quad \tilde
y=\frac{y}{H},\quad \tilde \eta =\frac{\eta}{a},\quad \tilde
t=\frac{t}{L/\sqrt{g H}},\quad \tilde \phi=\frac{\phi}{L
(a/H)\sqrt{g H}}\label{ND}
\end{equation}
where $g$ is the acceleration due to gravity, $H$ is the upstream
mean depth and $a$ and $L$ are typical values of the amplitude and
of the wavelength of the waves. Equations (\ref{6})--(\ref{9})
contain three non-dimensional parameters: the amplitude parameter
$\alpha=\frac{a}{H}$, the wavelength parameter
$\beta=\frac{H^2}{L^2}$ and the Bond number $\tau=\frac{T}{\rho g
H^2}$, where $T$ is the surface tension coefficient and $\rho$ is
the density of water.

Equations (\ref{6}) and (\ref{7})
are satisfied by making a standard substitution
\be\label{subsphi}
\phi=\sum\limits_{m=0}^{\infty}\frac{(-\beta)^m }{(2m)!}
\frac{\p^{2m} f(x,t)}{\p x^{2m}}y^{2m},
\ee
where
$f(x,t)=\phi|_{y=0}$. Substituting (\ref{subsphi}) into the
surface conditions (\ref{8}) and (\ref{9}) and differentiating
(\ref{9}) with respect to $x$ yields a system of two equations for
the surface elevation $\eta (x,t)$ and the horizontal velocity at
the bottom $w(x,t)=f_x$ in the form of infinite series with
respect to $\beta$.
We are interested in considering weakly nonlinear small amplitude
waves in a shallow water, so we will treat
$\alpha$ and $\beta$ as small parameters.

In the zero order in both $\alpha$ and $\beta$, the system of
equations for $w$ and $\eta$ reads $\eta_t+w_x=0, \; w_t+\eta_x=0$
so both $w$ and $\eta$ satisfy the linear wave equation
$\zeta_{tt}-\zeta_{xx}=0$ which describes waves traveling in two
directions. A wave moving to the right corresponds in this order
of approximation to $w=\eta$ and $\eta_t+\eta_x=0$. To derive the
equations describing right-moving waves in higher orders in
$\alpha$ and $\beta$, we can, along the lines of \cite{whitham},
reduce the system of equations for $w$ and $\eta$ to an
asymptotically equivalent set of equations consisting of a
relationship between the horizontal velocity $w$ and the surface
elevation $\eta$ and an evolution equation for the elevation. To
do this, we set \be\label{weq0}
w=\sum\limits_{i,j=0}^{\infty}R_{ij}\alpha^i\beta^j, \ee where
$R_{ij}$ depend on $\eta$ and its $x$-derivatives, and possibly
some nonlocal variables, with $R_{00}=\eta$, and require that
$\eta$ satisfy an evolution equation of the form \be\label{eteq0}
\eta_t=\sum\limits_{i,j=0}^{\infty}S_{ij}\alpha^i\beta^j, \ee
where $S_{00}=-\eta_x$ and in general $S_{ij}$ depend on $\eta$
and its $x$-derivatives. The functions $R_{ij}$ and $S_{ij}$ are
determined from the requirement of consistency of (\ref{weq0}) and
(\ref{eteq0}) with the above system of PDEs for $w$ and $\eta$. To
implement this, an iterative procedure starting from the zero
order of approximation and continuing to the higher orders is
applied, see the subsequent sections for details. In each order,
the $t$-derivatives of $\eta$ are replaced by their expressions
from the lower order equations.

However, it is obvious that
truncating the asymptotic expansions and keeping only the terms up
to certain order requires the knowledge of relationship between
the orders of magnitude of the two small parameters $\alpha$ and
$\beta$, because otherwise it is impossible to determine which
terms should be retained and which can be neglected. A commonly used
assumption is that
$\alpha$ and $\beta$
have the same order of magnitude ($\beta=O(\alpha)$). Then,
choosing, for example, $\alpha$ to be a primary parameter and
retaining the terms up to $O(\alpha^n)$,
we arrive at the so-called $n$-th order Boussinesq system
\cite{whitham}, \cite{bona}. If the first order Boussinesq system
is considered and the corresponding order expansions are taken for
(\ref{weq0}) and (\ref{eteq0}), then equation (\ref{eteq0})
for the elevation $\eta$ takes the form of the KdV equation
(\ref{01}), cf.\ \cite{whitham}, see also the next section. The same
procedure continued to next orders results in the KdV equation
with higher order corrections.
If the relationship $\beta=O(\alpha)$ does not hold,
then one needs an alternative assumption
relating the orders of $\alpha$ and $\beta$
to make a truncation of the expansions consistent.

\section{Procedure for the case of $\beta=O(\alpha)$}

In order to explain how the forms of the expansions (\ref{weq0})
and (\ref{eteq0}) are determined up to a certain order through an
iterative procedure, we will first present the procedure for the best
studied case of $\beta=O(\alpha)$ (we can set $\beta=\alpha$
without loss of generality).
We will also consider the problem without surface tension to make
the analysis as transparent as possible. Then equation (\ref{9})
is replaced by the following
\begin{eqnarray}
\label{9a} \phi_{t}+\frac{1}{2}\alpha \phi_{x}^2
+\frac{1}{2}\frac{\alpha}{\beta}\phi_{y}^2+\eta =0,\qquad
y=1+\alpha \eta.
\end{eqnarray}

If in the system of equations for $w$
and $\eta$ the terms in the second power of $\alpha$ are retained and the
higher order terms are dropped,
we arrive at the second order Boussinesq system
\begin{eqnarray}
\fl \eta_t+w_x+\alpha
\left(\left(w\eta\right)_x-\frac{1}{6}\eta_{3x}\right) +\alpha
^2\left(-\frac{1}{2}\left(\eta w_{2x}\right)_x+\frac{1}{120}w_{5x}\right)=0,\label{12}\\[3mm]
\fl w_t+\eta_x+\alpha \left(ww_x-\frac{1}{2}w_{2xt}\right)
+\alpha^2\left(-\left(\eta
w_{xt}\right)_x+\frac{1}{2}w_xw_{2x}-\frac{1}{2}w
w_{3x}+\frac{1}{24}w_{4xt}\right)=0.\label{13}
\end{eqnarray}

In the lowest (zero) order, the system (\ref{12}), (\ref{13}) reads
$\eta_t+w_x=0, \; w_t+\eta_x=0$, and the equivalent system
(\ref{weq0}), (\ref{eteq0}) describing a right-moving wave is
reduced to $w=\eta, \; \eta_t+\eta_x=0$. In the next order
iteration, we look for a solution for $w$ corrected to first order
as
\begin{equation}
\label{17} w=\eta +\alpha Q^{(1)},
\end{equation}
where $Q^{(1)}$ is a function of $\eta$ and its $x$-derivatives,
and substitute (\ref{17}) into equations (\ref{12}) and (\ref{13})
with the terms of order higher than $O(\alpha)$ dropped.  Upon the
substitution, the equations in question become
\begin{eqnarray}
\eta _t+\eta _x+\alpha \Bigl(2\eta \eta
_x-\frac{1}{6}\eta_{3x}+Q^{(1)}_x\Bigr)=0,\label{18a}\\[3mm]
\eta _t+\eta _x+\alpha \Bigl(2\eta \eta
_x-\frac{1}{2}\eta_{2xt}+Q^{(1)}_t\Bigr)=0.\label{18}
\end{eqnarray}
The function $Q^{(1)}$ is sought such that the two equations
(\ref{18a}) and (\ref{18}) agree (up to the first order in
$\alpha$) upon expressing all the $t$-derivatives of $\eta$ in
terms of its $x$-derivatives using the zero order
equation $\eta_t+\eta_x=0$. 
This yields $Q^{(1)}=-\frac{1}{4}\eta^2+\frac{1}{3}\eta _{2x}$.
Then equations (\ref{17}) and (\ref{18}) become
\begin{equation}
w=\eta+\alpha\Bigl(-\frac{1}{4}\eta^2+\frac{1}{3}\eta
_{2x}\Bigr),\quad \eta _t+\eta _x+\alpha \Bigl(\frac{3}{2}\eta
\eta _x+\frac{1}{6}\eta_{3x}\Bigr)=0.\label{21}
\end{equation}
The equation for $\eta$ is reduced to the KdV equation in a
standard form
\begin{equation}
\eta_{\hat t}+6\eta \eta_{\hat x}+\eta_{3\hat x}=0\label{KdVst}
\end{equation}
by the change of variables $(x,t)\rightarrow (\hat
x,\hat t)$, where
\begin{equation}
\hat x=\sqrt{\frac{3}{2}}\left(x-t\right),\quad \hat
t=\frac{1}{4}\sqrt{\frac{3}{2}}\;\alpha t.\label{KdVSc}
\end{equation}

At the next step, the above expression for $w$ is corrected to
second order in the form
\begin{equation}
\label{22} w=\eta+\alpha\Bigl( -\frac{1}{4}\eta^2+\frac{1}{3}\eta
_{2x}\Bigr)+\alpha^2 Q^{(2)}.
\end{equation}
It is substituted into (\ref{12}) and (\ref{13}), and then all the
$t$-derivatives of $\eta$ are replaced by their expressions
through the $x$ derivatives using the lower order equation,
namely, the second equation of (\ref{21}). The condition of
consistency of the two equations obtained in such a way leads to
an equation for $Q^{(2)}$, a solution of which is expressed in
terms of $\eta$ and its $x$ derivatives. As a result, we have
\begin{eqnarray}
\fl &w=\eta+\alpha\biggl( -\frac{1}{4}\eta^2+\frac{1}{3}\eta
_{2x}\biggr) +\alpha^2
\biggl(\frac{1}{8}\eta^3+\frac{3}{16}\eta_x^2
+\frac{1}{2}\eta\eta_{2x}
+\frac{1}{10}\eta_{4x}\biggr), \label{24}& \\[3mm]
\fl &\eta _t+\eta _x +\alpha
\biggl(\frac{3}{2}\eta\eta_x+\frac{1}{6}\eta_{3x}\biggr)
+\alpha^2\biggl(-\frac{3}{8}\eta^2\eta_x
+\frac{23}{24}\eta_x\eta_{2x} +\frac{5}{12}\eta\eta_{3x}
+\frac{19}{360}\eta_{5x}\biggr)=0.&\label{25}
\end{eqnarray}
The equation for $\eta$ can be reduced to the KdV equation with
the first order correction in standard form. The procedure for
determining the expansion (\ref{eteq0}) of the evolution equation
for the right-moving wave
can be continued to any order 
and yields the KdV equation with higher order corrections.

The next (third) order corrections with the terms up to the
seventh-order spatial derivatives included are given in
\cite{marchant}, \cite{bm}. The second and third order corrections
(up to seventh-order derivatives) for the case of nonzero surface
tension ($\tau\neq 0$) can be found in \cite{BurdeCNS} (at the end
of Appendix A) as a particular case of more general equations for
bi-directional waves. Note that in sections 4 (for $\tau=0$) and
5.3 (for $\tau\neq 0$) of the present paper corrections including
the terms up to the ninth-order spatial derivatives are calculated
for the case of $\alpha=O(\beta^2)$.

It should be emphasized once again that our procedure is aimed at
deriving equations for $\eta$ which have the form of an evolution
equation, and, accordingly, all the $t$-derivatives in the terms
of the order higher than zero are replaced by their expressions
through the $x$-derivatives. Therefore, applying this procedure
cannot yield equations which, like the BBM equation \cite{bbm},
contain the time derivatives in the higher order
terms.\looseness=-1

It is also worth noticing that there exists a possibility to
introduce certain freedom into the Boussinesq system. For example, a
class of Boussinesq systems which are formally equivalent to the
system displayed in (\ref{12})--(\ref{13}) can be derived
using other variables instead of the horizontal velocity at the
bottom $w$ and employing the lower order equations in
higher order terms \cite{whitham}, \cite{olver84}, \cite{bona}. It
might seem that this freedom, revealing itself as free
parameters present in the Boussinesq equations, should result in a
freedom in the equation for the surface elevation derived from
the Boussinesq equations under the assumption of
unidirectionality. However, this is not the case: it can be readily
checked that all those different but asymptotically equivalent
systems are reduced to the same high order KdV equation for the
surface elevation (\ref{25}) if the wave moving to the right is
specialized.

\section{Examples of ordering}

In this section, examples of a non-standard ordering when the
relation $\beta=O(\alpha)$ does not hold are studied. Two
important special cases, $\beta=O(\alpha^2)$ and
$\alpha=O(\beta^2)$, are considered in more detail. The analysis
is restricted to the pure gravity waves in order to better explain
the main points and demonstrate that the differences from the
standard model are not due to taking the surface tension into
account. The results for the problem including surface tension are
presented in the Appendix.

Starting from the case of $\beta=O(\alpha^2)$, we first write down
the system of equations for $w$ and $\eta$ obtained by keeping all
the terms of the order not higher than $\beta^2$, $\beta \alpha^2$
and $\alpha^4$:\looseness=-1
\begin{eqnarray}
&\eta_t+w_x+\alpha \left(\eta w\right)_x-\frac{1}{6}\beta w_{3x}
-\frac{1}{2}\alpha\beta \left(\eta w_{2x}\right)_x&\nonumber \\
&-\frac{1}{2}\alpha ^2\beta\left(\eta^2w_{2x}\right)_x
+\frac{1}{120}\beta^2 w_{5x}=0,&\label{m1}\\
&w_t+\eta_x+\alpha ww_x-\frac{1}{2}\beta w_{2xt} +\alpha
\beta\left(-\left(\eta
w_{xt}\right)_x+\frac{1}{2}w_xw_{2x}-\frac{1}{2}w
w_{3x}\right)&\nonumber \\
&+\alpha^2\beta\left(w_x\left(\eta
w_x\right)_x-\frac{1}{2}\left(\eta^2w_{xt}\right)_x-w\left(\eta
w_{2x}\right)_x
 \right)+\frac{1}{24}\beta^2 w_{4xt}=0.&\label{m2}
\end{eqnarray}
Next, we apply the iterative procedure described in the previous
section to determine the form of the unidirectional wave equations
(\ref{weq0}) and (\ref{eteq0}) for the case of
$\beta=O(\alpha^2)$. The resulting equations, with the terms up to
$O(\alpha^4)$ retained, read (we have used the square brackets to
gather the terms having the same order of magnitude)
\begin{eqnarray}
&w=\eta-\alpha\frac{\eta^2}{4}+\left[\alpha^2\frac{\eta^3}{8}+\beta\frac{\eta_{2x}}{3}\right]
+\left[-\alpha^3\frac{5\eta^4}{64}
+\alpha\beta\left(\frac{3\eta_x^2}{16}
+\frac{\eta\eta_{2x}}{2}\right)\right]&\nonumber\\
&+\left[\alpha^4\frac{7\eta^5}{128}
+\alpha^2\beta\left(\frac{\eta^2\eta_{2x}}{8}+\frac{3\eta\eta_x^2}{32}
+\frac{3z}{16}\right)
+\beta^2\frac{\eta_{4x}}{10}\right],\quad z=\int \eta_x^3 dx.&\label{Weq5}\\
&\eta_t+\eta_x+\frac{3}{2}\alpha\eta\eta_x+\left[-\frac{3}{8}\alpha^2\eta^2\eta_x
+\frac{1}{6}\beta\eta_{3x}\right]+\left[\frac{3}{16}\alpha^3\eta^3\eta_x
+\alpha\beta\left(\frac{23}{24}\eta_x\eta_{2x}+\frac{5}{12}\eta\eta_{3x}\right)\right]&
\nonumber\\
\label{Eteq5}&+\left[-\frac{15}{128}\alpha^4\eta^4\eta_x
+\alpha^2\beta\left(\frac{5}{16}\eta^2\eta_{3x}
+\frac{23}{16}\eta\eta_x\eta_{xx}+\frac{19}{32}\eta_x^3\right)
+\frac{19}{360}\beta^2\eta_{5x}\right]=0.& \label{defz}
\end{eqnarray}
To make the things even more clear, rewrite the last equation
taking $\beta=B \alpha^2$ ($B=O(1)$) and ordering the terms
according to powers of $\alpha$. We obtain
\begin{eqnarray}
&\eta_t+\eta_x+\frac{3}{2}\alpha\eta\eta_x+\alpha^2\left(-\frac{3}{8}\eta^2\eta_x
+\frac{1}{6}B\eta_{3x}\right)
+\alpha^3\left[\frac{3}{16}\eta^3\eta_x
+B\left(\frac{23}{24}\eta_x\eta_{2x}+\frac{5}{12}\eta\eta_{3x}\right)\right]&
\nonumber\\
&+\alpha^4\left[-\frac{15}{128}\eta^4\eta_x
+B\left(\frac{5}{16}\eta^2\eta_{3x}
+\frac{23}{16}\eta\eta_x\eta_{xx}+\frac{19}{32}\eta_x^3\right)
+\frac{19}{360}B^2\eta_{5x}\right]=0.&\label{Eteq5a}
\end{eqnarray}

It is immediate that an equation involving both nonlinearity and
dispersion is obtained at the leading order, which is now second
in $\alpha$ and first in $\beta$. Therefore this leading order
equation contains an extra term $-\frac{3}{8}\alpha^2\eta^2\eta_x$
and reads as follows:
\begin{equation}
\eta_t+\eta_x+\frac{3}{2}\alpha\eta\eta_x-\frac{3}{8}\alpha^2\eta^2\eta_x
+\frac{1}{6}\beta\eta_{3x}=0\label{Eteq6}.
\end{equation}
Thus, 
if $\beta=O(\alpha^2)$, then the leading order equation is not the
KdV equation but the Gardner equation which is a linear combination
of the KdV and of the modified KdV equation. The Gardner equation
has appeared in the literature in other physical contexts; in
particular, it was derived in an asymptotic theory for internal
waves in a two-layer liquid with a density jump at the interface
\cite{gard}, \cite{miles}. Our derivation shows that the Gardner
equation emerges in the classical water wave problem as the
leading order equation in the case of $\beta=O(\alpha^2)$. The
Gardner equation is integrable and possesses solitary wave
solutions but the Gardner solitons may differ in their properties
from their KdV counterparts, see e.g.\ \cite{pelin}.

The Gardner equation (\ref{Eteq6}) can be transformed into the
modified KdV equation
\begin{equation}
\tilde\eta_{\hat t}=\tilde\eta_{3{\hat
x}}+6\tilde\eta^2\tilde\eta_{\hat x},\label{MKdV}
\end{equation}
where $\tilde \eta$ is a shifted variable and $({\hat x},{\hat
t})$ are the rescaled variables in a moving frame. Equation
(\ref{MKdV}) is well known to be integrable, see e.g.\ \cite{as,
newell}. In addition to standard soliton solutions, it has
solutions in the form of `breather solitons' and also solutions
describing breather-soliton interactions. In view of the fact that
the transformation from (\ref{Eteq6}) to (\ref{MKdV}) includes a
shift of the dependent variable,  soliton solutions of equation
(\ref{MKdV}) for $\tilde\eta$ can be relevant for the original
problem in terms of $\eta$ if the flows with hydraulic jumps are
considered.

The form of the higher order corrections to the leading order
Gardner equation is also evident from equation (\ref{Eteq5}) (or
(\ref{Eteq5a})). Note that equation (\ref{Eteq5}) has  the
differential structure of a combination of the Gardner equation
(\ref{Eteq6})  and its first commuting flow; this feature is
similar to what is observed for the KdV with a higher order
correction in the case of $\beta=O(\alpha)$. \looseness=-1

In \cite{greki}, the so-called second and third order
approximations of water wave equations are studied for the case
$O(\beta)<O(\alpha)$ of \cite{fokas} specified to
$\beta\sim\alpha^2$. The comparison of these equations with
(\ref{Eteq5}) and (\ref{Eteq5a}) shows that the ordering (and
hence the truncation) used in \cite{greki} are invalid. In
particular, in the second-order approximation equation the terms
involving $\alpha\beta$ are present but the same order term
involving $\alpha^3$ is missing. Likewise, in the third-order
approximation equation the terms involving $\alpha^2\beta$ are
retained but the same order terms involving $\alpha^4$ and
$\beta^2$ are omitted.

Consider now the case $\alpha=O(\beta^2)$.
Then the basic system of equations for $w$ and $\eta$ obtained
by keeping the terms up to $O(\beta^4)$ (or $O(\alpha^2)$) has the form
\begin{eqnarray}
&\eta_t+w_x-\frac{1}{6}\beta w_{3x} +\alpha \left(\eta
w\right)_x+\frac{1}{120}\beta^2 w_{5x}
-\frac{1}{2}\alpha\beta\left(\eta w_{2x}\right)_x-\frac{1}{5040}\beta^3w_{7x}&\nonumber \\
&+\frac{1}{24}\alpha\beta^2\left(\eta
w_{4x}\right)_x+\frac{1}{362880}\beta^4w_{9x}
=0&\label{ab1}\\
&w_t+\eta_x-\frac{1}{2}\beta w_{2xt} +\alpha
ww_x+\frac{1}{24}\beta^2 w_{4xt}+\alpha\beta\left(-\left(\eta
w_{xt}\right)_x+\frac{1}{2}w_xw_{2x}-\frac{1}{2}w
w_{3x}\right)&\nonumber \\
&-\frac{1}{720}\beta^3w_{6xt}+\alpha\beta^2\left(\frac{1}{6}\left(\eta
w_{3xt}\right)_x+\frac{1}{12}w_{2x}w_{3x}
-\frac{1}{8}w_xw_{4x}+\frac{1}{24}w
w_{5x}\right)&\nonumber \\
&+\frac{1}{40320}\beta^4w_{8xt} =0&\label{ab2}
\end{eqnarray}
An equivalent system of the unidirectional wave equations
(\ref{weq0}) and (\ref{eteq0}) truncated to keep terms up to
$O(\beta^4)$ is (the meaning of the square brackets is the same as
in (\ref{Eteq5})):
\begin{eqnarray}
&w=\eta+\frac{1}{3}\beta
\eta_{2x}+\left[-\frac{1}{4}\alpha\eta^2+\frac{1}{10}\beta^2\eta_{4x}\right]
+\left[\frac{1}{16}\alpha\beta\left(3\eta_x^2+8\eta\eta_{2x}\right)
+\frac{61}{1890}\beta^3\eta_{6x}\right]&\nonumber\\
&+\left[\frac{1}{8}\alpha^2\eta^3+\alpha\beta^2\left(\frac{163}{360}\eta_{2x}^2
+\frac{1091}{1440}\eta_x\eta_{3x}+\frac{7}{20}\eta\eta_{4x}\right)
+\frac{1261}{113400}\beta^4\eta_{8x}\right]&\nonumber\label{Wab}\\
&\eta_t+\eta_x+\frac{1}{6}\beta\eta_{3x}+\left[\frac{3}{2}\alpha\eta\eta_x
+\frac{19}{360}\beta^2\eta_{5x}\right]
+\left[\alpha\beta\left(\frac{23}{24}\eta_x\eta_{2x}+\frac{5}{12}\eta\eta_{3x}\right)
+\frac{55}{3024}\beta^3\eta_{7x}\right]&\nonumber \\
&+\left[-\frac{3}{8}\alpha^2\eta^2\eta_x
+\alpha\beta^2\left(\frac{317}{288}\eta_{2x}\eta_{3x}
+\frac{1079}{1440}\eta_x\eta_{4x}+\frac{19}{80}\eta\eta_{5x}\right)
+\frac{11813}{1814400}\beta^4\eta_{9x}\right]
=0.&\nonumber\label{Etab}
\end{eqnarray}

It is immediate that for $\alpha=O(\beta^2)$ the equation
including at leading order both nonlinearity and dispersion is
\begin{equation}
\eta_t+\eta_x+\frac{1}{6}\beta\eta_{3x}+\frac{3}{2}\alpha\eta\eta_x
+\frac{19}{360}\beta^2\eta_{5x}=0\label{Etab5}
\end{equation}
By the change of variables
\begin{equation}
\hat x=\sqrt{\frac{3\alpha}{2\beta}}\left(x-t\right),\quad\hat
t=\frac{1}{4}\sqrt{\frac{3\alpha^3}{2\beta}}\;t
\end{equation}
equation (\ref{Etab5}) can be reduced to the following
\begin{equation}
\eta_{\hat t}+6\eta\eta_{\hat x}+\eta_{3\hat x} +M\eta_{5\hat
x}=0,\qquad M=\frac{19}{40}\alpha.\label{Etab5F}
\end{equation}
This equation, which is frequently referred to as the {\em 5th-order
KdV equation}, has been derived in \cite{hs} (with the parameter
$M$ defined in a different way) as a model equation for the
gravity-capillary shallow water waves of small amplitude when the
Bond number is close to but just less than $1/3$. It has been
extensively studied since then, see e.g.\ \cite{grma}, and,
although it is not integrable via the inverse scattering
transform, it is known to have a rich structure of solitary wave
solutions -- in particular, existence of nonlocal solitary
waves with propagating oscillatory tails and of asymmetric
solitary waves has been established. Our analysis shows that the
5th-order KdV equation (\ref{Etab5F}) arises as the leading order
equation in the classical water wave problem \textit{without
surface tension} when $\alpha=O(\beta^2)$.\looseness=-1

We will also present without derivation the leading order equation
for the case $\beta=O(\alpha^3)$ obtained by retaining the terms
which are at most cubic in $\alpha$. It reads
\begin{equation}
\eta_t+\eta_x+\frac32\alpha\eta\eta_x-\frac38\alpha^2\eta^2\eta_x
+\frac{3}{16}\alpha^3\eta^3\eta_x+\frac{\beta}{6}\eta_{3x}=0.
\end{equation}
This equation can be transformed into \be\label{k41a}
\eta_t=\eta^3\eta_x+\eta_{3x}, \ee which belongs to the type
$K(m,n)$ introduced by Rosenau and Hyman \cite{rh} with $m=4$ and
$n=1$. Equation (\ref{k41a})  is nonintegrable but admits
soliton-like traveling wave solutions in some range of wave
velocities.


\section{Discussion}

We have presented a procedure for systematic derivation of the
leading order and higher order evolution equations for the surface
elevation of unidirectional shallow water waves. This procedure is
based on a consistent ordering of terms in the original asymptotic
expansions for a prescribed relationship between orders of
magnitude of two small parameters $\alpha$ and $\beta$. Our
results provide a set of consistent model equations for
unidirectional water waves which replace the KdV equation and the
higher order KdV equations in the cases when the parameters
$\alpha$ and $\beta$ are not of the same order of magnitude. Some
of the equations emerging in our analysis as the leading order
equations in the asymptotic expansion for the unidirectional water
waves have been proposed before as model equations in other
physical contexts (e.g., the Gardner equation, the modified KdV
equation, and the so-called 5th-order KdV equation).
In the higher orders of approximation, a variety of evolution
equations which can serve as higher order models for
unidirectional water waves on equal footing with the higher order
KdV equations are found. Our analysis also reveals that certain
model equations used in the literature are questionable since they
have been obtained as a result of an improper ordering which is
invalid for any relationship among orders of $\alpha$ and $\beta$.



\noindent The present analysis is based on assuming a prescribed
relationship between orders of magnitude of two small parameters
$\alpha$ and $\beta$.
However, the results can be interpreted in another, alternative
way, along the lines of the analysis presented in \cite{cj}.
The main concern of the analysis of \cite{cj} is to demonstrate
that the condition $\beta=O(\alpha)$ is not necessary for having a
balance between nonlinearity and dispersion characteristic of the
KdV equation and that the KdV balance is possible for any $\beta$
provided that $\alpha \rightarrow 0$. To this end the variables
are transformed in such a way that the parameter $\beta$ is scaled
out, in favour of $\alpha$, which leads to a prescription, in
asymptotic terms, of the region of time and space where the
balance occurs and so the KdV equation is valid. This conceptual
shift from a relationship between orders of magnitude of the two
small parameters to distances and times needed for achieving the
balance between nonlinearity and dispersion provides a new view
which is more relevant to applications in nature.

However, the analysis of \cite{cj} is restrictive  in the sense
that the transformation of variables introduced in \cite{cj} may
result only in the problem which leads to the KdV equation to
leading order as $\alpha\rightarrow 0$. In what follows, we show
that it is not because of some intrinsic properties of the water
wave equations but simply due to a specific character of the
transformation used in \cite{cj}. We extend the analysis of
\cite{cj} by introducing a generalized transformation dependent on
a parameter $n$  (the transformation of \cite{cj} becomes a
particular case). This generalized transformation, like the
transformation introduced in \cite{cj}, results in the system of
equations  which contains only one small parameter $\alpha$.
Specifying the transformation parameter $n$ to different values
allows to obtain a variety of different problems and a variety of
the corresponding leading order equations (like the Gardner
equation, the 5th-order KdV equation and so on) including the KdV
equation. As a matter of fact, each problem obtained from the
original one by applying the transformation for a particular value
of $n$ can be equivalently obtained by assuming the relationship
$\beta=O(\alpha^n)$ between orders of magnitude of the small
parameters. The former approach allows to consider the
nonlinear-dispersion balance, epitomized by the soliton equations,
as existing for any $\beta$, provided that $\alpha \rightarrow 0$,
but imposes conditions on the regions of space and time in which
the soliton dynamics (the KdV dynamics, the Gardner dynamics, the
5th-order KdV dynamics and so on) are expected to occur.

In \cite{cj}, the transformations eliminating $\beta$ are applied
to the original system of equations in terms of velocities
$(u,w)$, pressure $p$ and elevation $\eta$ and then the system of
equations in terms of $\phi$ and $\eta$ is obtained from the
transformed equations. Therefore, in our analysis, we will also
deal with the original equations (although the same could be done
for equations (\ref{6})--(\ref{9}) in terms of $\phi$ and
$\eta$). The system of equations of a two-dimensional irrotational
wave, with effects of surface tension negligible, after
non-dimensionalising takes the form
\begin{eqnarray}
&&u_t+\alpha \left(u u_x+w u_y\right)=-p_x,\qquad
\beta\left(w_t+\alpha \left(u w_x+w w_y\right)\right)=-p_y,\label{D1}\\
&&u_x+w_y=0,\qquad u_y-\beta w_x=0,\label{D2}\\
&&w=0\qquad \mathrm{on}\quad y=0,\\
&&p=\eta,\quad w=\eta_t+\alpha u \eta_x\qquad \mathrm{on}\quad
y=1+\alpha \eta.\label{D3}
\end{eqnarray}
(In the notation of \cite{cj}, $y\rightarrow z,\;\beta\rightarrow
\delta^2,\; \alpha\rightarrow \epsilon$.) The scales for $(x,y,t)$
are as in (\ref{ND}) and the scales for $u$, $w$ and $p$ are
respectively $(a/H)\sqrt{g H}$, $(a/L)\sqrt{g H}$ and $\rho g a$.

The following transformations are applied to equations (\ref{D1})
-- (\ref{D3}) in \cite{cj}:
\begin{eqnarray}
&&x\rightarrow \frac{\sqrt{\beta}}{\sqrt{\alpha}}\;x,\quad
y\rightarrow y,\quad t\rightarrow
\frac{\sqrt{\beta}}{\sqrt{\alpha}}\;t,\qquad p\rightarrow p,\quad
\eta\rightarrow \eta,\quad w\rightarrow
\frac{\sqrt{\alpha}}{\sqrt{\beta}}\;w\label{trcj}
\end{eqnarray}
As the result the system (\ref{D1})--(\ref{D3}) reduces to the
system of equations
\begin{eqnarray}
&&u_t+\alpha \left(u u_x+w u_y\right)=-p_x,\qquad
\alpha\left(w_t+\alpha \left(u w_x+w w_y\right)\right)=-p_y,\label{D1a}\\
&&u_x+w_y=0,\qquad u_y-\alpha w_x=0,\label{D2a}\\
&&w=0\qquad \mathrm{on}\quad y=0,\\
&&p=\eta,\quad w=\eta_t+\alpha u \eta_x\qquad \mathrm{on}\quad
y=1+\alpha \eta\label{D3a}
\end{eqnarray}
which are the same as (\ref{D1})--(\ref{D3}), but with $\beta$
replaced by $\alpha$, for arbitrary $\beta$.
From the analysis made in section 3 of the present paper (and from
an equivalent analysis of \cite{cj}) it is evident that equations
(\ref{D1a})--(\ref{D3a}) constitute the representation that
leads to the KdV equation (\ref{KdVst}) to leading order as
$\alpha\rightarrow 0$.

As it was explained above, the transformations (\ref{trcj}) can be
generalized. The generalized transformations are
\begin{eqnarray}
&&x\rightarrow \frac{\sqrt{\beta}}{\alpha^{n/2}}\;x,\quad
y\rightarrow y,\quad t\rightarrow
\frac{\sqrt{\beta}}{\alpha^{n/2}}\;t,\qquad p\rightarrow p,\quad
\eta\rightarrow \eta,\quad w\rightarrow
\frac{\alpha^{n/2}}{\sqrt{\beta}}\;w\label{trcjG}
\end{eqnarray}
where $n$ is arbitrary. Applying the transformations (\ref{trcjG})
to equations (\ref{D1})--(\ref{D3}) results in the system
\begin{eqnarray}
&&u_t+\alpha \left(u u_x+w u_y\right)=-p_x,\qquad
\alpha^n\left(w_t+\alpha \left(u w_x+w w_y\right)\right)=-p_y,\label{D1b}\\
&&u_x+w_y=0,\qquad u_y-\alpha^n w_x=0,\label{D2b}\\
&&w=0\qquad \mathrm{on}\quad y=0,\\
&&p=\eta,\quad w=\eta_t+\alpha u \eta_x\qquad \mathrm{on}\quad
y=1+\alpha \eta\label{D3b}
\end{eqnarray}
which is (\ref{D1})--(\ref{D3}), with $\beta$ replaced by
$\alpha^n$.

It is clear that the same problem (\ref{D1b})--(\ref{D3b})
results also from the assumption $\beta=O(\alpha^n)$ which allows
to replace $\beta$ by $\alpha^n$ without loss of generality. As a
matter of fact, all the results of the present paper are related
to the problem (\ref{D1b})--(\ref{D3b}), independently of the
approach through which it is obtained. The analysis made in
Sections 3 and 4 indicates that equations (\ref{D1b})--(\ref{D3b}) 
specified to different values of $n$ give rise to
different equations to leading order as $\alpha\rightarrow 0$, The
KdV equation arises as a particular case for $n=1$. Other
particular cases might be the Gardner equation ($n=2$), the
5th-order KdV equation ($n=1/2$),  the $K(4,1)$-type
 equation in the sense of \cite{rh} ($n=3$). In the case, when the system  
(\ref{D1b})--(\ref{D3b}) is treated as obtained via the transformations
(\ref{trcjG}), the results are valid under some conditions on the
regions of space and time where thus the corresponding soliton
dynamics are expected to occur. It should be emphasized, however,
that, although the system (\ref{D1b})--(\ref{D3b}) can be
equivalently obtained either by applying the transformations
(\ref{trcjG}) or by assuming $\beta=O(\alpha^n)$, these two
approaches represent alternative views which cannot be combined.

\ack The research of the second author was supported by the
postdoctoral fellowship at the Jacob Blaustein Institutes for
Desert Research of the Ben-Gurion University of the Negev and by
the Grant Agency of the Czech Republic under grant P201/12/G028.
The authors thank the referees for useful suggestions.

\section*{Appendix. Results for the case of nonzero surface tension}

Since the multiplier $(1-3\tau)$ appears in the coefficients of
the highest derivatives in the leading order equations, the case
of $\tau= 1/3$ should be considered separately. We will assume
that $\tau\neq 1/3$ and moreover, that $|\tau-1/3|$ is {\em not}
small. Indeed, if $|\tau-1/3|\ll 1$, one has to introduce yet
another small parameter $\epsilon=\tau-1/3$ and consider the
asymptotic expansion with respect to $\epsilon$ as well, see e.g.\
\cite{hs}.

\subsection{$\beta=O(\alpha^2)$}

We will consider the case of $\beta=O(\alpha^2)$ keeping the terms
that are at most quartic in $\alpha$. Following the procedure
described above, we obtain
\begin{eqnarray}
&w=\eta-\frac{\alpha\eta^2}{4}+\frac{\alpha^2\eta^3}{8}+\frac{\beta}{6}(2-3\tau)\eta_{2x}
-\frac{5}{64}\alpha^3\eta^4&\nonumber\\
&+\alpha\beta\left(\left(3+7\tau\right)\frac{\eta_x^2}{16}
+4\left(2+\tau\right)\eta\eta_{2x}\right)
&\nonumber\\
&+\frac{7\alpha^4\eta^5}{128}
+\frac{\alpha^2\beta}{32}\left(2(2-3\tau)\eta^2\eta_{2x}+3(1-7\tau)\eta\eta_x^2
+6(1-\tau)z\right)
&\nonumber\\
&-\frac{\beta^2}{120}(-12 + 20 \tau + 15 \tau^2)\eta_{4x},&\label{weq5}\\
&\eta_t+\eta_x+\frac{3}{2}\alpha\eta\eta_x-\frac{3}{8}\alpha^2\eta^2\eta_x
+\frac{\beta}{6}(1-3\tau)\eta_{3x}+\frac{3}{16}\alpha^3\eta^3\eta_x &\nonumber\\
&+\frac{\alpha\beta}{24}\left((23+15\tau)\eta_x\eta_{2x}+2(5-3\tau)\eta\eta_{3x}
\right)
-\frac{15}{128}\alpha^4\eta^4\eta_x&\nonumber\\
&+\frac{\alpha^2\beta}{32}\left(2(5+\tau)\eta^2\eta_{3x}
+2(23-5\tau)\eta\eta_x\eta_{2x}
+(19-13\tau)\eta_x^3\right)&\nonumber\\
&-\frac{\beta^2}{360}(-19 + 30 \tau + 45\tau^2)\eta_{5x}
=0,&\label{eteq5}
\end{eqnarray}
where $z=\int \eta_x^3 dx$.

If we keep in (\ref{eteq5}) the terms of order not greater than
$O(\alpha^2)$ to retain the dispersion and nonlinearity at the
leading order then it reads
\begin{equation}
\label{Gard} \eta_t+\eta_x
+\frac32\alpha\eta\eta_x-\frac38\alpha^2\eta^2\eta_x+\frac{\beta}{6}(1-3\tau)\eta_{3x}=0.
\end{equation}
This is nothing but the Gardner equation which for $\tau=0$
coincides with equation (\ref{Eteq6}) discussed in Section 4. This
equation can be reduced to the modified Korteweg--de Vries
equation
$$
\eta_t=K\eta_{3x}+6\eta^2\eta_x,
$$
where $K=\mathop{\rm sign}\nolimits (\beta(1-3\tau))$.

\subsection{$\beta=O(\alpha^3)$}

Keeping the terms up to the order of $\alpha^5$ we have

\begin{eqnarray}
\label{eteq6}&w=\eta-\frac{\alpha\eta^2}{4}+\frac{\alpha^2
\eta^3}{8}
-\frac{5\alpha^3\eta^4}{64} +\frac{\beta(2-3\tau)}{6}\eta_{2x}&\nonumber\\
&+\frac{7\alpha^4\eta^5}{128}
+\frac{\alpha\beta}{4}\left((2+\tau)\eta\eta_{2x}
+(3+7\tau)\frac{\eta_x^2}{4}\right)&\nonumber\\
&-\frac{21\alpha^5\eta^6}{512}+\frac{\alpha^2\beta}{16}\left(3(7\tau-1)\frac{\eta\eta_x^2}{2}
-3(\tau-1)z+(2-3\tau)^2\eta^2\eta_{2x}\right)&
\\
&\eta_t+\eta_x+\frac32\alpha\eta\eta_x-\frac{3}{8}\alpha^2\eta^2\eta_x
+\frac{3}{16}\alpha^3\eta^3\eta_x+\frac{\beta(1-3\tau)}{6}\eta_{3x}&\nonumber\\
&-\frac{15}{128}\alpha^4\eta^4\eta_x+\frac{\alpha\beta}{24}((23+15\tau)\eta_x\eta_{2x}
-2(3\tau-5)\eta\eta_{3x})+\frac{21}{256}\alpha^5\eta^5\eta_x&\nonumber\\
&+\frac{\alpha^2\beta}{32}((-13\tau+19)\eta_x^3
+2(23-5\tau)\eta\eta_x\eta_{2x}+2(\tau+5)\eta^2\eta_{3x})=0,&
\end{eqnarray}
where again $z=\int \eta_x^3 dx$.

If we consider the leading order equation, i.e., restrict
ourselves to the terms which are at most cubic in $\alpha$, then
(\ref{eteq6}) becomes
\begin{equation}
\eta_t+\eta_x+\frac32\alpha\eta\eta_x-\frac38\alpha^2\eta^2\eta_x
+\frac{3}{16}\alpha^3\eta^3\eta_x+\frac{\beta}{6}(1-3\tau)\eta_{3x}=0.
\end{equation}
This equation can be further transformed into \be\label{k41}
\eta_t=M\eta^3\eta_x+\eta_{3x}, \ee where $M=\mathop{\rm
sign}\nolimits(\beta(1-3\tau))$. Eq.(\ref{k41}) belongs to the
type $K(m,n)$ introduced by Rosenau and Hyman \cite{rh} with $m=4$
and $n=1$. It is nonintegrable but it is readily seen to admit
soliton-like traveling wave solutions for $M=1$ in a certain range
of wave velocities.

\subsection{$\alpha=O(\beta^2)$}

In this case we have

\begin{eqnarray}
&w=\eta+\frac{\beta}{6}(2-3\tau)\eta_{2x}+\beta^2\bigl(\frac{1}{10}-\frac{\tau}{6}-\frac{\tau^2}{8}\bigr)\eta_{4x}
-\frac{\alpha\eta^2}{4}&\nonumber\\
&+\beta^3(488 - 756 \tau -630 \tau^2 - 945
\tau^3)\frac{\eta_{6x}}{15120}
+\alpha\beta\left((3+7\tau)\frac{\eta_x^2}{16}+(2+\tau)\frac{\eta\eta_{2x}}{4}\right)&\nonumber\\
&+\beta^4\left(\frac{1261}{113400} - 61 \frac{\tau}{3780} -
\frac{\tau^2}{80} - \frac{\tau^3}{48}
- \frac{5 \tau^4}{128}\right)\eta_{8x}&\nonumber\\
&+\alpha\beta^2\biggl((326 + 435 \tau + 315
\tau^2)\frac{\eta_{2x}^2}{720} +(28 - 20 \tau
+ 5 \tau^2)\frac{\eta \eta_{4x}}{80}&\nonumber\\
&+(1091 + 480 \tau + 945 \tau^2)\frac{\eta_x \eta_{3x}}{1440}\biggr) +\frac{\alpha^2\eta^3}{8},&\label{e1ab2}\\
&\eta_t+\eta_x+\frac{\beta}{6}(1-3\tau)\eta_{3x}+\frac32\alpha\eta\eta_x
+\frac{\beta^2}{360}(19-30 \tau-45 \tau^2)\eta_{5x}&\nonumber\\
&+\frac{\alpha\beta}{24}\left((23+15
\tau)\eta_x\eta_{2x}+2(5-3\tau)\eta\eta_{3x}\right)
&\nonumber\\
&-\frac{\beta^3}{15120}(-275+399 \tau+315 \tau^2+945
\tau^3)\eta_{7x}
&\nonumber\\
&-\frac38\alpha^2\eta^2\eta_x
+\alpha\beta^2
\biggl((317+270 \tau+441
\tau^2)\frac{\eta_{2x}\eta_{3x}}{288}+\left(1079-150\tau+
855\tau^2\right)\frac{\eta_x\eta_{4x}}{1440}&\nonumber\\
&-(-57+50 \tau+15
\tau^2)\frac{\eta\eta_{5x}}{240}\biggr)&\nonumber\\
&+\beta^4\biggl(\frac{11813}{1814400}-\frac{55
\tau}{6048}-\frac{19
\tau^2}{2880}-\frac{\tau^3}{96}-\frac{5\tau^4}{128}\biggr)\eta_{9x}=0.&\label{e2ab2}
\end{eqnarray}

Notice that in the second order in $\beta$, which is the leading
order in this case, the equation (\ref{e2ab2}) for $\eta$ can be
transformed into the Korteweg--de Vries equation only if $19-30
\tau-45 \tau^2=0$ ($\tau\approx 0.4$) when the term with
$\eta_{5x}$ vanishes. If $19-30 \tau-45 \tau^2\neq 0$, then
(\ref{e2ab2}) in second order in $\beta$ reads
$$
\eta_t+\eta_x+\frac{\beta}{6}(1-3\tau)\eta_{3x}+\frac{3}{2}\alpha\eta\eta_x
+\frac{\beta^2}{360}(19-30 \tau-45 \tau^2)\eta_{5x}=0.
$$
We can get rid of the term $\eta_x$ by passing from $x$  to
$x'=x-t$, so upon omitting the prime at $x$ the equation under
study becomes
$$
\eta_t+\frac{\beta}{6}(1-3\tau)\eta_{3x}+\frac{3}{2}\alpha\eta\eta_x
+\frac{\beta^2}{360}(19-30 \tau-45 \tau^2)\eta_{5x}=0.
$$
Next, let $x=A \tilde x$, $t=B \tilde t$, $\eta=C\tilde \eta$,
where $B=360A^5/(\beta^2(-19+30 \tau+45 \tau^2))$,
$C=-\beta^2(-19+30\tau+45 \tau^2)/(1080\alpha A^4)$. Then upon
omitting tildes at $x,t,\eta$ we obtain the so-called 5th-order
KdV equation (see Section 4) in the form \be\label{eteqnonint1}
\eta_t=\eta\eta_x+K\eta_{3x}+\eta_{5x}, \ee where $K=(3\tau-1)60
A^2/(\beta(-19+30 \tau+45 \tau^2))$. Assuming $\tau\neq 1/3$, we
can set $A=(|(\beta(-19+30 \tau+45 \tau^2)/(60(3\tau-1))|)^{1/2}$,
and then $K=\mathop{\rm sign}\nolimits((\beta(-19+30 \tau+45
\tau^2)/(3\tau-1))$.

\subsection{$\alpha=O(\beta^3)$}

If $\alpha=O(\beta^3)$, we obtain
\begin{eqnarray}
&w=\eta+\frac{\beta}{6}(2-3 \tau)\eta_{2x}
+\beta^2 \left(\frac{1}{10}-\frac{\tau}{6}-\frac{\tau^2}{8}\right)\eta_{4x}&\nonumber\\
&-\frac{\alpha\eta^2}{4}
-\frac{\beta^3}{15120}\left(-488 + 756 \tau + 630 \tau^2 + 945 \tau^3\right)\eta_{6x}&\nonumber\\
&+\beta^4\left(\frac{1261}{113400} - \frac{61 \tau}{3780} -
\frac{\tau^2}{80} - \frac{\tau^3}{48} - \frac{5
\tau^4}{128}\right) \eta_{8x}
+\frac{\alpha\beta}{16}((3+7 \tau) \eta_{x}^2+4(2+\tau)\eta \eta_{2x})&\nonumber\\
&-\frac{\beta^5}{39916800}\Bigl(-159264 + 221936 \tau + 161040
\tau^2
+ 249480 \tau^3&\nonumber\\
&+ 519750 \tau^4 +1091475 \tau^5\Bigr)\eta_{10x}
+\frac{\alpha\beta^2}{1440}\Bigl(18(28 - 20 \tau + 5
\tau^2)\eta\eta_{4x}&\nonumber\\
&+(1091 + 480 \tau + 945 \tau^2)\eta_x\eta_{3x} +2(326 + 435 \tau + 315 \tau^2)\eta_{2x}^2\Bigr),&\label{eqw3}\\
&\eta_t+\eta_x+\frac{\beta}{6}(1-3\tau)\eta_{3x}
+\frac{\beta^2}{360}(19-30 \tau-45 \tau^2)\eta_{5x}&\nonumber\\
&+\frac32\alpha\eta\eta_x-\frac{\beta^3}{15120}\biggl(-275+399 \tau+315 \tau^2+945 \tau^3\biggr)\eta_{7x}&\nonumber\\
&+\beta^4\biggl(\frac{11813}{1814400}-\frac{55}{6048}\tau
-\frac{19}{2880}\tau^2-\frac{\tau^3}{96}-\frac{5\tau^4}{128}\biggr)\eta_{9x}&\nonumber\\
&+\frac{\alpha\beta}{24}\left(2(3\tau-5)\eta\eta_{3x}+(23+15
\tau)\eta_x\eta_{2x}\right)
&\nonumber\\
&+\frac{\beta^5}{39916800}\left(95265 - 129943 \tau - 90750 \tau^2
-131670 \tau^3 - 259875 \tau^4 -1091475 \tau^5\right)\eta_{11x}&\nonumber\\
&+\alpha\beta^2\biggl((317+270 \tau+441
\tau^2)\frac{\eta_{2x}\eta_{3x}}{288}
+(1079-150 \tau + 855 \tau^2)\frac{\eta_x\eta_{4x}}{1440}&\nonumber\\
&-(-57 + 50 \tau + 15 \tau^2)\frac{\eta\eta_{5x}}{240}\biggr).&
\label{eqeta3}
\end{eqnarray}
In this case the leading order is three, and upon introducing a
new variable $x'=x-t$ the leading order equation for $\eta$
reads
\begin{eqnarray}
&\eta_t+{\ds\frac{\beta}{6}}(1-3\tau)\eta_{3x}
+{\ds\frac{\beta^2}{360}}(19-30 \tau-45 \tau^2)\eta_{5x}\nonumber\\
&+{\ds\frac{3\alpha}{2}}\eta\eta_x-{\ds\frac{\beta^3}{15120}}(-275+399
\tau+315 \tau^2+945 \tau^3)\eta_{7x}=0.
\end{eqnarray}


\section*{References}

\begin{thebibliography}{99}

\bibitem{whitham}Whitham G B
1974 \textit{Linear and Nonlinear Waves} (Wiley/Interscience, New
York)

\bibitem{as}Ablowitz M and Segur H 1981
\textit{Solitons and the inverse scattering transform (SIAM
Studies in Applied Mathematics, 4)} (Society for Industrial and
Applied Mathematics (SIAM): Philadelphia, PA)

\bibitem{newell}Newell A C  1985 \textit{Solitons in mathematics and
physics (CBMS-NSF Regional Conference Series in Applied
Mathematics, 48)} (Society for Industrial and Applied Mathematics
(SIAM), Philadelphia, PA)

\bibitem{johnson1}Johnson R S 2003 The Classical Problem of Water Waves: a Reservoir of
Integrable and Nearly-Integrable Equations \textit{J. Nonlin.
Math. Phys.} \textbf{10} Suppl. 1 72-92

\bibitem{gottwald0}Dullin H R, Gottwald G A and Holm D D 2001 An
integrable shallow water equation with linear and nonlinear
dispersion \textit{Phys. Rev. Lett.} \textbf{87} 4501-4

\bibitem{greki}Tzirtzilakis E, Marinakis V, Apokis C and Bountis T
2002 Soliton-like solutions of higher order wave equations of the
Korteweg--de Vries type \textit{J. Math. Phys.} \textbf{43}
6151-65

\bibitem{kichenassamy} Kichenassamy S and Olver P J 1992 Existence and nonexistence
of solitary wave solutions to higher-order model evolution
equations \textit{SIAM J. Math. Anal.} \textbf{23} 1141-66

\bibitem{burde1} Burde G I 2010 Generalized Kaup--Kupershmidt solitons and other
solitary wave solutions of the higher order KdV equations
\textit{J. Phys. A: Math. Theor.} \textbf{43} 085208 (13pp)

\bibitem{burde2} Burde G I 2011 Static algebraic solitons in
Korteweg--de Vries type systems and the Hirota transformation
\textit{Phys. Rev. E} \textbf{84} 026615 (11pp)


\bibitem{bbm}Benjamin T B,  Bona J L and Mahoney J J 1972
Model equations for long waves in nonlinear dispersive systems
\textit{Phil. Trans. Roy. Soc. Lond.} \textbf{A 272} 47-78

\bibitem{bona} Bona J L,  Cnen M and Saut J C 2002 Boussinesq
equations and other systems for small mplitude long waves in
nonlinear dispersive media. I. Derivation and linear theory
\textit{J. Nonlin. Sci.} \textbf{12} 283-318

\bibitem{marchant} Marchant T R 2002 High-order interaction of solitary waves on shallow water
\textit{Studies in Applied Math.} {\bf 109} 1-17

\bibitem{bm}Bhatt R and Mikhailov A V 2010 On the inconsistency of the Camassa-Holm model
with the shallow water theory {\it Preprint} arXiv:1010.1932v1

\bibitem{BurdeCNS} Burde G I 2011 Solitary wave solutions of the high-order KdV models
for bi-directional water waves \textit{Commun. Nonlinear Sci.
Numer. Simulat.} \textbf{16} (2011) 1314-–28

\bibitem{olver84} Olver P J 1984 Hamiltonian and non-Hamiltonian models for water
waves \textit{Trends and Applications of Pure Mathematics in
Mechanics (P.G. Ciarlet and M. Roseau, eds) (Lecture Notes in
Physics, No. 195)} (Springer-Verlag, New York) 273-90

\bibitem{gard}Kakutani T and Yamasaki T 1978 Solitary waves on a
two-layer fluid \textit{J. Phys. Soc. Jpn.} \textbf{45} 674-79

\bibitem{miles}Miles J W  1979 On internal solitary waves
\textit{Tellus} \textbf{31} 456-62

\bibitem{pelin}Pelinovski\u{\i} E N and Slyunyaev A V 1998 Generation and interaction of lsrge-amplitude solitons
\textit{JETP Letters} \textbf{67} 655-61

\bibitem{fokas}Fokas A S 1995 On a class of physically important
integrable equations \textit{Physica D} \textbf{87} 145-50

\bibitem{hs}Hunter J K and Scheurle J 1988 Existence of perturbed solitary wave solutions to a model
equation for water waves \textit{Phys. D} \textbf{32} 253-68

\bibitem{grma}Grimshaw R, Malomed B and Benilov E 1994 Solitary
waves with damped oscillatory tails -- an analysis of the
5th-order Korteweg--de Vries equation \textit{Physica D}
\textbf{77} 473-85

\bibitem{rh}Rosenau P and Hyman J M 1993 Compactons: Solitons with finite wavelength
\textit{Phys. Rev. Lett.} \textbf{70} 564-67

\bibitem{cj}Constantin A and Johnson R S 2008 On the non-nimensionalisation,
scaling and resulting interpretation of the classical governing
equations for water waves \textit{J. Nonlin. Math. Phys.}
\textbf{15} Supplement 2 58-73







\end{thebibliography}
\end{document}